\documentclass[AMA,STIX1COL]{WileyNJD-v2}

\articletype{Article Type}%

\received{26 April 2016}
\revised{6 June 2016}
\accepted{6 June 2016}

\usepackage{float}

\begin{document}

\title{Bayesian local exchangeability design for phase II basket trials}

\author[1]{Yilin Liu}

\author[1]{Michael Kane}

\author[1]{Denise Esserman}

\author[1]{Ondrej Blaha}

\author[1]{Daniel Zelterman}

\author[1]{Wei Wei*}


\address[1]{\orgdiv{Department of Biostatistics}, \orgname{Yale School of Public Health}, \orgaddress{\state{New Haven, Connecticut 06520}}}



\corres{*Wei Wei, Department of Biostatistics, Yale School of Public Health, New Haven, CT 06520. \email{wei.wei@yale.edu}}


\abstract{We propose an information borrowing strategy for 
the design and monitoring of phase II basket trials based on the local multisource exchangeability assumption between baskets (disease types). In our proposed local-MEM framework,  information borrowing is only allowed to occur locally, i.e., among baskets with similar response rate and the amount of information borrowing is determined by the level of similarity in response rate, whereas baskets not considered similar are not allowed to share information. We construct a two-stage design for phase II basket trials using the proposed strategy. The proposed method is compared to competing Bayesian methods and Simon's two-stage design in a variety of simulation scenarios. We demonstrate the proposed method is able to maintain the family-wise type I error rate at a reasonable level and has desirable basket-wise power compared to Simon's two-stage design.  In addition, our method is computationally efficient compared to existing Bayesian methods in that the posterior profiles of interest can be derived explicitly without the need for sampling algorithms. R scripts to implement the proposed method are available at https://github.com/yilinyl/Bayesian-localMEM.}

\keywords{master protocol, oncology, adaptive design, clustering, Bayesian analysis}


\maketitle


\section{Introduction}
Breakthroughs in molecular biology have led to the development of more personalized treatment strategies targeting specific molecular aberrations involved in tumor growth. As a molecular aberration can occur in tumors of different histological or anatomical types, the traditional "one indication at a time" strategy evaluating a new treatment in a certain cancer (sub)type is no longer sustainable. To accelerate the oncology drug development process, there has been a growing interest in conducting basket trials, which provide a framework for simultaneously testing the anti-tumor activity of a novel agent in a variety of cancer (sub)types harboring the same therapeutic target.\cite{woodcock2017master, redig2015basket, renfro2017statistical}. In a basket trial, the term basket or indication refers to a cohort of patients with the same cancer (sub)type. Patients in different baskets are commonly treated by the same targeted agent. The majority of proof of concept, phase II basket trials were conducted as independent parallel studies without a concurrent control arm\cite{mcneil2015nci, middleton2020national}.

Consider a prototypical basket trial based on the Vemurafenib study. The Vemurafenib trial was designed to test the preliminary efficacy of Vemurafenib, a BRAF inhibitor, in BRAFV600 mutation positive non-melanoma cancers in six pre-specified baskets comprising NSCLC, cholangiocarcinoma (bile duct or BD), Erdheim-Chester disease or Langerhans' cell histiocytosis (ED.LH), anaplastic thyroid cancer (ATC), and colorectal cancer (CRC). 
The statistical challenges in implementing a basket trial can be described based on this prototype.  
A pooled analysis combining data across baskets may be performed if we assume the treatment effect is homogeneous across different baskets. This assumption is often invalid and the effect of a treatment can be significantly different across baskets. For example, Vemurafenib was found to be effective in treating BRAF V600E mutant melanoma and hairy cell leukemia, but not for BRAF mutant colon cancer. \cite{flaherty2010inhibition, tiacci2011braf, prahallad2012unresponsiveness} When the homogeneity assumption is invalid, then pooled analysis will lead to biased estimate of the treatment effect. Alternatively, we may perform basket-wise analysis. But such analysis often suffer from a lack of statistical power owing to the difficulty of accruing patients from rare disease (sub)types and the sample size constraints of most phase II studies. In the Vemurafenib study, the ATC and BD basket only enrolled 7 and 8 patients, respectively, as compared to the 26 patients in the CRC basket.

A variety of statistical strategies have been proposed to improve the accuracy and efficiency of basket trials. \cite{thall2003hierarchical, liu2017increasing, berry2013bayesian, chu2018bayesian, chu2018blast, hobbs2018bayesian, neuenschwander2016robust, simon2016bayesian, zhou2020robot, kang2021hierarchical} The Bayesian hierarchical model approach (BHM) formulated by Thall et al.\cite{thall2003hierarchical} and Berry et al.\cite{berry2013bayesian} aims to improve the efficiency of basket trials by enabling information borrowing across baskets. The validity of the BHM relies on the assumption of single source exchangeability (SSE), which assumes the response rates from different baskets arise from a common parent distribution defined by trial-level parameters. However, in practice a basket trial often consists of mixtures of treatment responsive and treatment resistant subtypes. Under these circumstances,  the effectiveness of the experimental treatment cannot be adequately characterized by a unimodal distribution defined based on the SSE assumption and the traditional BHM approach often underestimate the heterogeneity between baskets. It has been shown BHM based on the SSE assumption can lead to inappropriate borrowing in the presence of non-exchangeable baskets, resulting in inflated type I error rate and reduced power\cite{hobbs2018bayesian}.

To address the limitations of the traditional BHM approach, Neuenschwander \cite{neuenschwander2016robust} and Berry \cite{simon2016bayesian} proposed different methods to measure exchangeability between baskets.
Hobbs et al. \cite{hobbs2018bayesian} developed a basket trial design based on the multisource-exchangeability model (MEM), which allows the identification of exchangeable baskets as well as singleton baskets. Zhou et al. \cite{zhou2020robot} infers the number of subgroups and subgroup memberships using a Dirichlet process mixture model. Kang et al. propose a hierarchical Bayesian clustering design that clusters arms into either active or inactive subgroup \cite{kang2021hierarchical}. These aforementioned approaches are computationally demanding and are difficult to implement due to the complexities involved in model calibration and prior specifications. 

We  propose  an  information  borrowing  strategy under the local multisource exchangeability assumption and construct a two-stage design for phase II basket trials using the proposed strategy. The proposed work differs from previous approaches in the following aspects. First, to address the pooling vs. not pooling issue, we formally evaluate the evidence for and against the heterogeneity of treatment effects 
using Bayes factor and information sharing is only allowable if there is sufficient evidence for pooling some baskets together. Second, to determine the extent as well as the amount of information sharing, the proposed approach partitions baskets into mutually non-exchangeable blocks and conducts local information borrowing within each block according to the similarity of response profiles for baskets in the same block. We implement the proposed information borrowing strategy in both interim and final analysis such that non-promising baskets can be removed earlier from the trial. Third, compared to all the existing approaches, our method is easy to implement because we don't require multiple tuning parameters for model calibration and all the posterior quantities of interest can be derived without the need for sampling algorithms. 

In Section \ref{sec:pool}, we consider whether or not to allow data sharing across baskets by formulating a set of hypotheses representing different levels of between-basket heterogeneity. Section \ref{sec:similarity} then describes our method for conducting posterior inferences under the local multisource exchangeability assumption based on these hypotheses.  We construct stopping boundaries for making go/no go decisions in Section \ref{sec:decision}. We show results from simulation studies under varying response scenarios in Section \ref{sec:sim_setup} based on comparisons with MEM and Simon two-stage design. Finally, we discuss limitations and future directions in Section \ref{discussion}.

\section{Methods} \label{method}
\subsection{To borrow or not to borrow} \label{sec:pool}
Consider a phase II basket trial consisting of $B$ baskets, each with a binary endpoint indicating treatment success or failure. We model each basket as a sequence of independent Bernoulli samples with success probability $\theta_b$ for $b=1,\ldots, B$. Denote $x_b$ as the number of responses out of the $n_b$ patients in basket $b$. The number of successful responses in a basket is assumed to have a binomial distribution, 
$$x_b\sim Binomial(n_b, \theta_b).$$

A crucial decision in the design and analysis of basket trials is whether data can be borrowed from different baskets. We will evaluate whether data should be borrowed and to what extent it should be borrowed by setting up $J$ hypotheses representing all possible partitions of the 
$B$ baskets into blocks of varying sizes. 
We assume baskets within the same block have identical response rate, whereas baskets in different blocks respond differently to the experimental treatment.  Under this setup, we will only consider information sharing for baskets from the same block and we will not permit information sharing for baskets in different blocks. 

For illustration purpose, consider $B=4$ baskets, which can form $J=15$ different partitions (Table \ref{tab:pp}). The hypothesis/partition $\boldsymbol\Omega_1$ assumes the response rates for all the $B$ baskets are identical 
$$\boldsymbol\Omega_1:\, \theta_{1}=\cdots=\theta_{4}.$$
We should conduct pooled analysis under $\boldsymbol\Omega_1$.

Hypotheses $\Omega_2, \cdots, \Omega_{J-1}$ represent different local borrowing scenarios when
information sharing is limited to baskets within the same block. Specifically, we consider
$$\boldsymbol\Omega_2:\, \theta_{1}=\theta_{2}=\theta_3\neq\theta_{4}$$
$$\boldsymbol\Omega_3:\, \theta_{1}=\theta_2=\theta_{4}\neq\theta_{3}$$
$$\cdots,$$
$$\boldsymbol\Omega_{J-1}:\, \theta_{1}\neq\theta_{2}\neq\theta_{3}=\theta_4,$$
and the hypothesis represents the scenario where each basket has its unique response rate and the differences of response rate between baskets are clinically meaningful. That is, the $B$ baskets belong to $B$ non-exchangeable blocks
$$\boldsymbol\Omega_J:\, \theta_{1} \ne \theta_{2} \ne \theta_{3}\ne \theta_{4}.$$
Under $\mathbf{\Omega_J}$, basket-wise analysis should be conducted and information sharing between baskets is not permitted.

We assume there are $K_j$ blocks under $\boldsymbol\Omega_j$, and each block has its unique response rate $\theta_{jk}$, where $k=1, \cdots, K_j$. For example, we have $K_1=1$, $K_2=2$, $K_3=2$ unique response rates under $\boldsymbol\Omega_1$, $\boldsymbol\Omega_2$ and $\boldsymbol\Omega_3$, respectively. We denote the block membership of the basket $b$ under $\boldsymbol\Omega_j$ as $\omega_{jb}$. 

The plausibility of each hypothesis can be estimated based on the observed response rates of the $B$ baskets. Let $S_{jk}$ and $N_{jk}$ denote the sum of responses and sample sizes for baskets in the $k$-th block of $\boldsymbol\Omega_j$, then
$$
S_{jk} | \theta_{jk} \sim \text{Binomial}(N_{jk}, \theta_{jk}).
$$
Let $beta$ denote the probability density function of a Beta distribution. 
We assume $\theta_{jk}$ has a non-informative prior and follows the Beta distribution 
with probability density function $\pi(\theta_{jk})$, where
 $$ \pi(\theta_{jk}) = beta(\alpha_0, \beta_0).
 $$
Common choices of $\pi(\theta_{jk})$ include $beta(1,1)$ and $beta(0.5, 0.5)$.
 
The posterior distribution of $\theta_{jk}$ under the hypothesis $\mathbf{\Omega_j}$ is
\begin{equation}
\pi(\theta_{jk}|S_{jk}, N_{jk}, \mathbf{\Omega_j}) = beta(\alpha_0+S_{jk}, \beta_0+N_{jk}-S_{jk}),
\end{equation}

Let $\mathcal{B}$ denote the Beta function. Under $\mathbf{\Omega_j}$, the marginal density of $S_{jk}$ is 
$$m(S_{jk}|\mathbf{\Omega_j})=\left(\!\begin{array}{c}
     N_{jk}  \\
     S_{jk} 
\end{array}\!\right) \frac{\mathcal{B}(\alpha_0+S_{jk}, \beta_0+N_{jk}-S_{jk})}{\mathcal{B}(\alpha_0, \beta_0)}.$$

With the use of Bayes's rule, the posterior probability of $\mathbf{\Omega_j}$ is 
\begin{equation}
\pi(\mathbf{\Omega_j}|\mathbf{S_j}, \mathbf{N_j})= \frac{\prod_k {m(S_{jk}|\mathbf{\Omega_j)}} \pi(\mathbf{\Omega_j})}{\sum_j \prod_k {m(S_{jk}|\mathbf{\Omega_j)}}\pi(\mathbf{\Omega_j})},
\end{equation}
where $\mathbf{S_j}=(S_{j1}, \cdots, S_{jK_j})$ and $\mathbf{N_j}=(N_{j1}, \cdots, N_{jK_j})$.

These $J$ hypotheses represent different decisions we have to make in the analysis of basket trials. Let $\Omega^*$ denote the partition with the largest posterior probability. The decision to borrow or not to borrow and where to borrow will be based on $\Omega^*$. If $\Omega^*=\Omega_J$, we will conduct basket-wise analysis. If $\Omega^*=\Omega_1$, we will consider information sharing among all the $B$ baskets. We will only consider local borrowing for other choices of $\Omega^*$.

\subsection{Prior Specification}\label{prior}
We consider the following prior specifications for the these $J$ hypotheses:
$$\pi(\mathbf{\Omega_j})=\frac{K_j^{\delta}}{\sum_{j=1}^J K_j^{\delta}},$$
for $j=1, \cdots, J.$

When $\delta=0$, we have $\pi(\mathbf{\Omega_j})=1/J$ and we consider the $J$ hypotheses as equally likely in a priori. When $\delta>0$, this prior favors partitions with more blocks. That is, partitions representing higher level of between-basket heterogeneity are given more weight and we becomes more reluctant to borrow as $\delta$ increases.
When $\delta<0$, partitions favoring more pooling are given more prior weight. We investigate the effect of different prior choices by considering $\delta=0, 1, 2$. 

We enumerate all the possible partitions of four baskets in Table \ref{tab:pp}. Table \ref{tab:pp} also shows different prior probabilities for these partitions assuming $\delta=0, 1, 2$. 

\subsection{A local borrowing strategy} \label{sec:similarity}
Specifying the $J$ hypotheses not only helps us to carefully weigh the decision of borrowing vs not pooling, it also provides a framework for the evaluation of pairwise similarity between baskets. We define pairwise similarity as the posterior probability of two baskets residing in the same block, i.e., having the same response rate. Specifically, the pairwise similarity between the $s$-th and $t$-th basket is 
\begin{equation}
\label{omega}
\Psi_{st}=\sum_{j=1}^J \pi(\mathbf{\Omega_j}|\mathbf{S_j}, \mathbf{N_j})I(\omega_{js}=\omega_{jt}),
\end{equation}
for $s\neq t$. By definition, there is $\Psi_{st}=\Psi_{ts}$. We consider $\Psi_{st}=1$ if $s=t$. 

The pairwise similarity between two baskets is calculated by summing up the posterior probabilities of partitions which include these two baskets in the same block. A pairwise similarity of 1/0 indicates two baskets are fully exchangeable/non-exchangeable, whereas a similarity between 0 and 1 suggests two baskets are only partially exchangeable. 

Denote $\boldsymbol{\Psi}$ as the symmetric $B\times B$ similarity matrix with elements $\Psi_{st}$ defined by equation (\ref{omega}). Previous works in MEM assume information sharing can occur between any two baskets and the amount of information sharing is proportional to the pairwise similarity between baskets.\cite{kaizer2019basket, kaizer2021statistical, kane2019analyzing, hobbs2018bayesian} We refer to these approaches as "global-MEM". Under global-MEM,  the posterior distribution for the response probability of basket $b$ is 
\begin{equation}
\label{posterior}
\theta_b|\mathbf{x}, \mathbf{n}, \boldsymbol\Psi \sim \text{Beta}\left\{\alpha_0 + \sum_{t=1}^B\Psi_{bt}x_t, \, \beta_0 + \sum_{t=1}^B\Psi_{bt}y_t\right\},
\end{equation}
where $\mathbf{x}=(x_1, \cdots, x_B)$,  $\mathbf{n}=(n_1, \cdots, n_B)$ and $y_t=n_t-x_t$.

The use of pairwise similarity in global-MEM can lead to excessive borrowing as it cannot accurately capture the uncertainty of simultaneously borrowing from multiple baskets. Following our local borrowing strategy, the decision to borrow or not to borrow and where to borrow is determined by the top partition $\Omega^*$. Under $\Omega^*$, information is shared locally for baskets within the same block. The posterior probability of $\Omega^*$ denoted as $\pi(\Omega^*)$ evaluates the uncertainty related to the local borrowing decision. Unlike previous MEM approaches, the posterior distribution for the response probability of basket $b$ in the proposed approach is defined as
\begin{equation}
\label{posterior}
\theta_b|\mathbf{x}, \mathbf{n}\sim \text{Beta}\left\{\alpha_0 + x_b + \pi(\Omega^*)\sum_{t\neq b}x_tI(\omega^*_{b}=\omega^*_{t}),\, \beta_0 + y_b + \pi(\Omega^*)\sum_{t\neq b}y_tI(\omega^*_b=\omega^*_t)\right\}.
\end{equation}

Basket-level inference relies on $\Omega^*$ and its posterior probability $\pi(\Omega^*)$. First, information sharing is conducted at a local scope with the boundaries set by $\Omega^*$. A basket is only allowed to borrow from a different basket if two baskets are estimated to have the same block membership ($\omega^*_b=\omega^*_t$) under $\Omega^*$. Second, when borrowing is allowed, the amount of information borrowing is proportional to $\pi(\Omega^*)$. The more certainty we have about our local borrowing decision, the more data we borrow from neighboring baskets. Hereafter, we will refer to this  approach as "local-MEM". 

In the Bayesian context, the amount of information sharing across baskets is measured by the effective sample size (ESS) of the resultant posterior distribution.\cite{hobbs2013adaptive, kaizer2018bayesian} The ESS of basket $b$ based on equation (\ref{posterior}) is
\begin{equation}
\label{ess}
ESS_b=\alpha_0+\beta_0+ n_b + \pi(\Omega^*)\sum_{t\neq b} n_tI(\omega^*_b=\omega^*_t).
\end{equation}

When $\Omega^*=\Omega_J$,  the effective sample size of basket $b$ is reduced to
$$ESS_b=\alpha_0+\beta_0+ n_b,$$
which is the effective sample size if basket-wise analysis is performed. 

\subsection{Go/no go decisions for monitoring a basket trial} \label{sec:decision}

In exploratory basket trials, the primary objective is to evaluate whether the experimental treatment is worthy of further investigation for each basket. Denote $\theta_{b0}$ as the fixed, pre-specified response rate seen in historical controls for basket $b$ and let $\theta_{b1}$ be the target response rate of clinical interest. This primary objective is accomplished by testing the null hypothesis
$$H_{b0}:\, \theta_b\leq\theta_{b0}$$ 
versus the alternative
$$H_{b1}:\, \theta_b>\theta_{b0}, $$ 
evaluating the statistical power at $\theta_{b1}$, where $\theta_{b1}>\theta_{b0}$.

Consider a two-stage design allowing early termination for futility. Denote $N_b$ as the pre-specified maximum sample size for basket $b$. Let $n_{b}$ denote the pre-specified sample size for basket $b$ at stage I. We consider $n_{b}=tN_b$, where $0<t<1$. Let $q_1$ and $q_2$ denote the posterior probability cutoff at the first and second stage. Similar to Zhou et al.,  \cite{zhou2017bop2} we define the stopping boundary based on a two-parameter power function such that 
$q_1=\lambda t^{\gamma}$ and $q_2=\lambda$, where $\lambda$ and $\gamma$ can be tuned to provide desirable operating characteristics. With the use of the two-parameter power function, our design can be extended to accommodate multiple interim analyses. In practice, we can choose $t$ based on the projected accrual rate and the maximum sample size for each basket. We recommend setting $t=0.5$ such that the interim analysis is conducted when 50\% of information has been accrued in each basket. 

Let $ P_{b} = Pr(\theta_b>\theta_{b0}|\mathbf{x}, \mathbf{n})$ denote the posterior probability of having a higher response rate than $\theta_{b0}$ in basket $b$ based on our local borrowing strategy. We define the following rules for making go/no go decisions:

\begin{itemize}
    \item At stage I, if $ P_{b} \le q_{1}$, then terminate accrual for futility for basket $b$, 
    \item otherwise, continue basket $b$ to the next stage.
    \item At stage II, update $\Omega^*$ and $P_{b}$ using data from the $B'$ baskets remaining in the trial. Due to possible early termination of baskets, we have $B'\leq B$. 
    \item If $ P_{b} >  q_{2}$, then conclude the new treatment is promising for this basket. 
    \item We conclude the new treatment is not promising in basket $b$ if $ P_{b} \leq  q_{2}$
\end{itemize}

\begin{figure}[ht]
\centering
\includegraphics[scale=0.8]{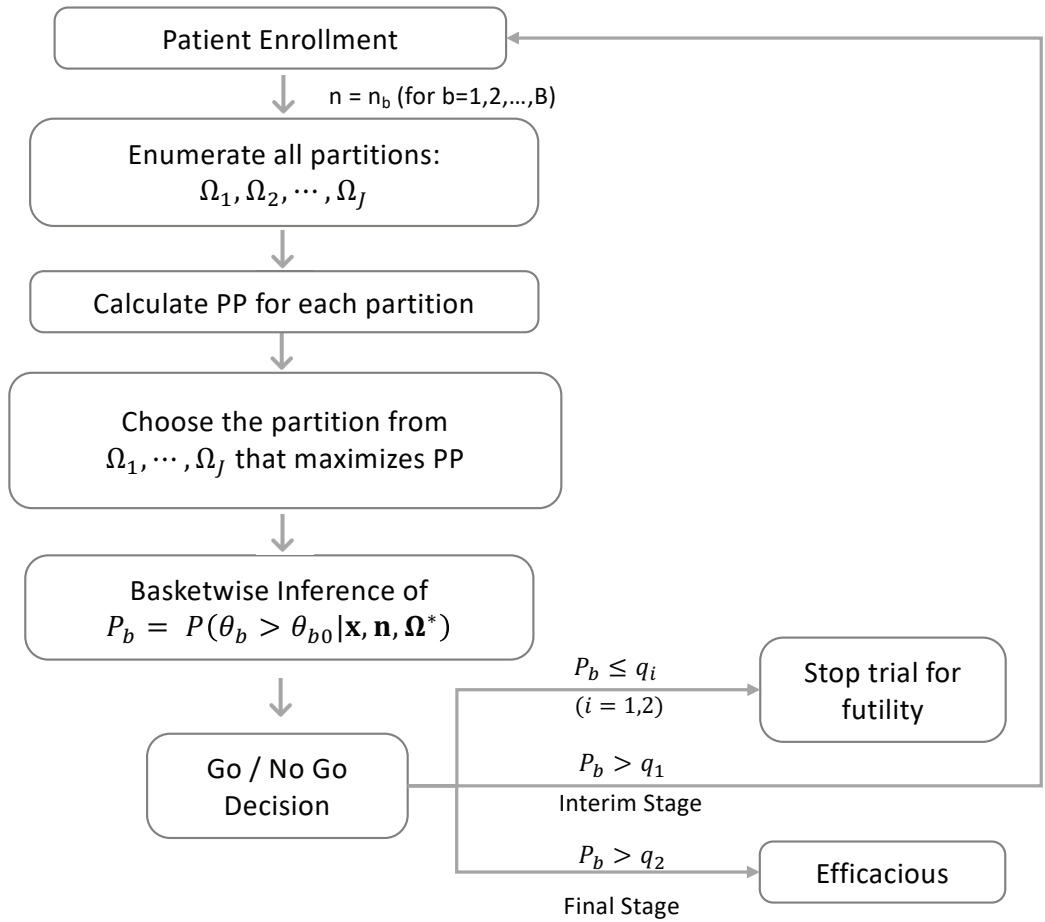}
\caption{Flowchart of the two-stage design based on the local-MEM method. Denote $\mathbf{x}, \mathbf{n}$ as vectors for the number of responses and pre-specified sample sizes for all baskets, $q_i$ as the posterior stopping boundary at stage $i$. \label{flowchart}}
\end{figure}

Figure \ref{flowchart} shows the workflow of our proposed design.

The frequentist operating characteristics of the proposed approach can be evaluated based on the following criteria via simulation studies: 
\begin{itemize}
    \item Family-wise type I error rate (FWER): This the probability that the treatment is wrongly claimed to be efficacious for at least one non-promising basket.
    \item Basket-wise type I error rate: Consider a single basket $b$. This is the probability of rejecting the null hypothesis when the treatment is actually non-promising. 
    \item Basket-wise power: Consider a single basket $b$. This is the probability of rejecting the null hypothesis when the treatment is actually efficacious.
    \item Trial-wise power: this is defined as the weighted average of basket-wise power for all promising baskets, where the weight is proportional to the sample size of each promising basket.
\end{itemize}

In practice, the maximum sample size $N_b$ of each basket is often determined by budget or accrue rate. Given the maximum basket size $N_b$, our choice of the decision boundary is the pair ($q_1, q_2$) which maximizes the trial-wise power of the design under the global alternative, i.e., all baskets have promising response rates ($\theta_b=\theta_{b1}$), while controlling the FWER at a target level (e.g.,0.1) under the global null, i,e, the treatment is not effective in any baskets($\theta_b=\theta_{b0}$).  

\section{Simulation} \label{simulation}

\subsection{Simulation setup and evaluation metrics} \label{sec:sim_setup}

We conducted simulation studies to evaluate the operating characteristics of the proposed design under different response scenarios.
Consider a total of four baskets ($b=1, \ldots, 4$).  The response scenarios we considered include the global null ("0 success" or "4 failures"), the global alternative ("4 success"), and the mixed alternative scenarios. Under the global null scenario, the response rate in each basket is the same as the historical control rate ($\theta_{b0} = 0.15$) and we should not declare success in any of these baskets. Under the global alternative, the response rate is promising ($\theta_{b1} = 0.45$) in all the baskets. The mixed alternative scenarios consist of baskets with both promising ($\theta_{b1} = 0.45$) and non-promising ($\theta_{b0} = 0.15$) response rates, representing varying levels of between-basket heterogeneity.  

We evaluated these scenarios here on two specific cases:

\begin{itemize}
    \item {\bf{Case 1}} A fixed design ($N_b=19$) is used for each basket.
    
    \item {\bf{Case 2}} A two-stage design is applied to each basket, assuming the same maximum sample size ($N_b=16$) for all the baskets.
\end{itemize}

 We simulate 5,000 trials for each response scenario under each case. 
 In case 1, we evaluate the performance of the proposed local-MEM method, the global-MEM,\cite{hobbs2018bayesian, zabor2022bayesian} the EXNEX approach\cite{neuenschwander2016robust}, the RoBoT approach\cite{zhou2020robot} and basket-wise analysis based on basket-wise power, basket-wise type I error rate and FWER under each scenario. Hobbs et al.\cite{hobbs2018bayesian} considered a prior pairwise exchangeability of 0.50 in MEM assuming impartiality in terms of exchangeability between different baskets in priori, and Zabor et al.\cite{zabor2022bayesian} recommended setting the prior probability of pairwise exchangeability between 0.1 and 0.3 with the aim of providing a better trade-off between power and FWER control. Therefore, we implemented the global-MEM approach assuming the prior probability of pairwise exchangeability is 0.50 and 0.10, respectively. We refer to these different MEM implementations as global-MEM (0.5) and global-MEM (0.1), respectively. Alternatively, Neuenschwander et al. proposed the exchangeability-nonexchangeability (EXNEX) approach  \cite{neuenschwander2016robust} which performs information borrowing by allowing each basket-specific parameter to be either exchangeable with other similar baskets or nonexchangeable with any of them. We implement EXNEX following the recommendations of Neuenschwander et al. \cite{neuenschwander2016robust}. Specifically, we model the response rate in each basket as a mixture of two exchangeable (EX) distributions and one non-exchangeable (NEX) distribution. Weights for the EX and NEX components are set to 0.25, 0.25, and 0.5, respectively. Prior distributions are chosen for the EX parameters such that the variance of these priors correspond to approximately one observation, whereas the NEX parameters are fixed. Posterior computation with RoBoT is based on 10,000 MCMC samples following 50,000 iterations of burn-in. Following Zhou and Ji, \cite{zhou2020robot}
 We assume $\alpha=2$, which determines the prior probability for the number of latent subgroups. We also conduct basket-wise analysis assuming a uniform prior for the response rate of each basket. All these approaches are calibrated to maximize the trial-wise power under global alternative while maintaining the FWER under the global null scenario at a level less than 0.10. In case 2, we compared the local-MEM method with Simon's two-stage design on the basis of basket-wise power, basket-wise type I error rate, FWER and expected sample size. Likewise, the FWER under the global null scenario is maintained at comparable levels ($< 0.10$) for the proposed method and the Simon two-stage design. The code for simulation is written in R\cite{team2013r}. The scripts can be found at https://github.com/yilinyl/Bayesian-localMEM. The comparison results will be discussed in the following sections.

We implement the global-MEM method using the `basket' package \cite{RJ-2021-020} in R. The EXNEX approach is implemented using the R package `bhmbasket'. The R code for RoBoT is obtained from \url{https://github.com/tianjianzhou/RoBoT}.

\subsection{Simulation results under fixed design} \label{sec:mem_cmp}

\begin{figure}[ht]
\centering
\includegraphics[scale=0.8]{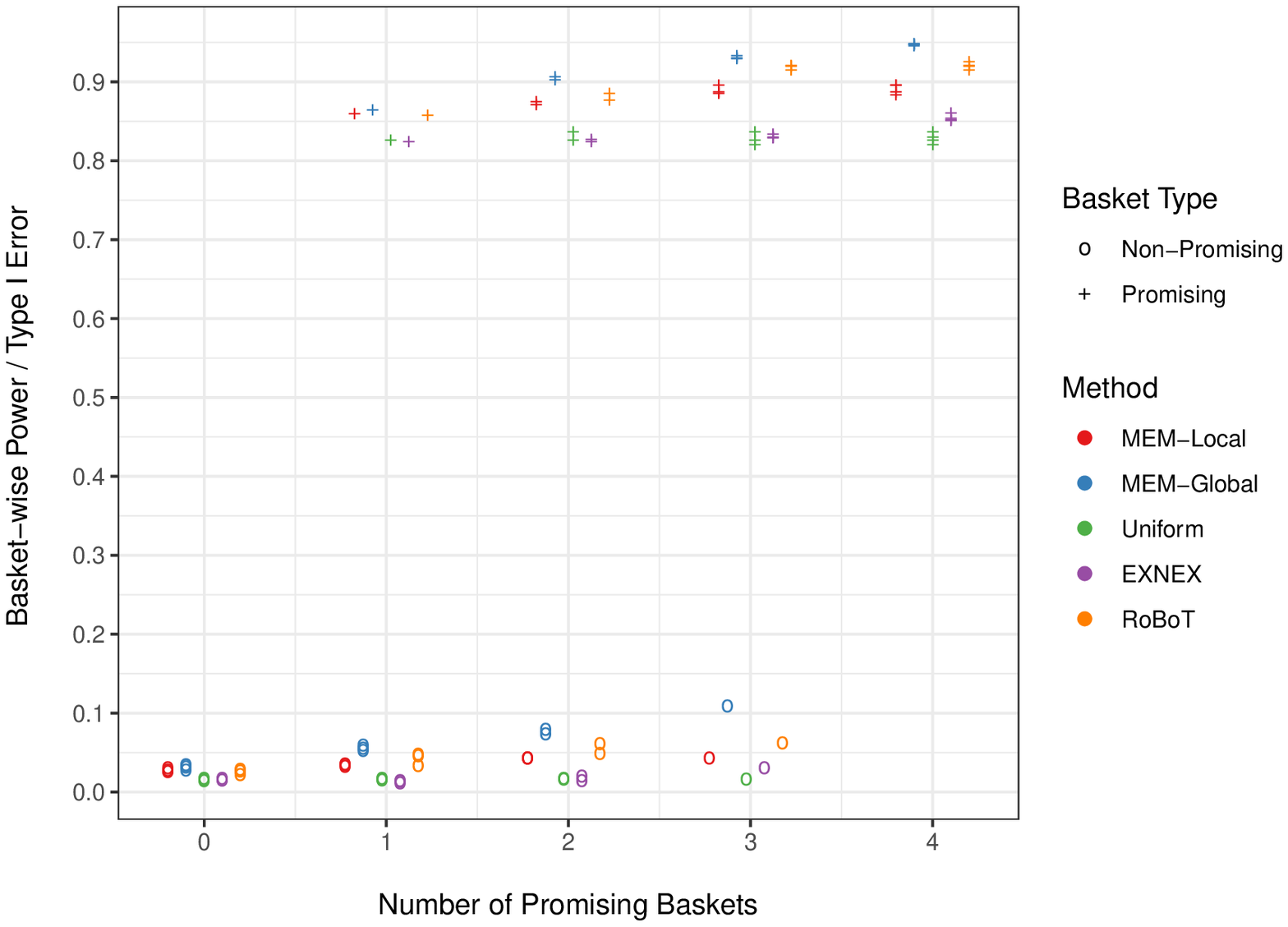}
\caption{Local-MEM ($\delta=2$) vs. global MEM (prior exchangeability $=0.1$) vs. basket-wise analysis with uniform prior vs. EXNEX vs. RoBoT for studies with fixed design. Each basket has a fixed sample size of 19 without provisions for early stopping. Different methods are color coded. FWER under the global null scenario is controlled at 0.10 level for each method. \label{B4_homo_cmp}}
\end{figure}

In case 1, all baskets have a fixed sample size of $N_b=19$. Figure \ref{B4_homo_cmp} compares the performance of local-MEM($\delta=2)$ with global-MEM (0.1), EXNEX, RoBoT and basket-wise analysis using uniform prior under case 1. Basket-wise power (or basket-wise type I error rate when a basket is not promising) is displayed for each basket. As shown in Figure \ref{B4_homo_cmp}, the EXNEX and basket-wise analysis with uniform prior have similar levels of basket-wise power and type I error rates among different scenarios, except that EXNEX achieves greater power ($>85\%$) than basket-wise analysis ($82\%-84\%$) when all baskets are promising. Both local- and global-MEM achieve greater basket-wise power than EXNEX and basket-wise analysis.  The minimum basket-wise power across all scenarios using the local-MEM($\delta=2$) method is $86\%$, which is achieved when only a single basket is promising ("1 success"). This is comparable to the basket-wise power of global-MEM(0.1) and RoBoT under the same scenario and is noticeably higher than the power of EXNEX and basket-wise analysis. When at least three baskets are truly promising, the local-MEM($\delta=2$) method achieves a basket-wise power of close to $90\%$. As more effective baskets are added to the mixed alternative scenarios, the global-MEM approach tends to borrow more information than other approaches, showing an increasing trend in both basket-wise power and type I error rate. Although the global-MEM(0.1) model achieves greater power than the local-MEM($\delta=2$) method when there are at least "2 success" out of four baskets, it also leads to inflated type I error rate in the non-promising baskets. In particular, when the treatment is truly promising in all but one basket ("3 success"), the type I error rate for the non-promising basket exceeds $0.10$. By contrast, the basket-wise type I error rate using the local-MEM($\delta=2$) method is $< 0.05$ under all scenarios.

Table \ref{tab:b4_compare} provides more detailed comparisons for all the previously mentioned methods under different simulation scenarios. This table also shows the impact of different prior choice, i.e., $\delta$ on local-MEM. As $\delta$ increases, partitions with more blocks (less information borrowing) are assigned with higher probability in priori. As shown in Table \ref{tab:b4_compare}, basket-wise power achieved by local-MEM decreases as $\delta$ increases from 0 to 2. The local-MEM with $\delta=0$ achieves the greatest basket-wise power among all local-MEM models, but it leads to an modest inflation in FWER in the "1 success" (FWER$=0.134$) and "2 success" (FWER$=0.116$) scenarios.
The local-MEM with $\delta=1$ shows better control of FWER compared to $\delta=0$, with a FWER of 0.117 and 0.104 under the "1 success" and "2 success" scenarios. By contrast, the local-MEM with $\delta=2$ keeps both family-wise and basket-wise type I error rate under control, while providing power gains compared to the basket-wise analysis and the EXNEX approach. Regardless the choice of prior, the basket-wise type I error rate of local-MEM is under 0.10 in all scenarios. The global-MEM (0.1 and 0.5) achieve greatest basket-wise power in most scenarios. However, Both global-MEM (0.1) and global-MEM (0.5) have inflated FWER especially when the prior probability of pairwise exchangeability is set to 0.50. When there are 3 promising baskets and one single nonpromising basket ("3 success"), the global-MEM(0.5) has a FWER of 0.218. The RoBoT approach maintains the FWER at reasonable levels in all scenarios. Under the "1 success" scenario, it has a FWER of 0.124 and has FWER levels close to 0.10 under the other scenarios. The performance of RoBoT is similar to the local-MEM approach with $\delta=0$ and $\delta=1$. 


To summarize, in our 4-basket simulation setting, the local-MEM approach based on different prior choices provides improved statistical power without overly inflating the FWER compared to basket-wise analysis.  If basket-wise type I error rate control is desired, we recommend setting $\delta=0$ or $\delta=1$ for the local-MEM. If more stringent control of FWER is needed, we recommend the local-MEM with a larger $\delta$. Compared to the global-MEM method, the local-MEM approach is less likely to borrow from the "wrong" baskets because information sharing is limited to a similarity defined local scope,  and maintains basket-wise type I error rates at reasonable levels in all scenarios. Compared to the EXNEX method, the local-MEM provides more statistical power and is easy to implement without the need for intensive model calibration. Compared to RoBoT, the local-MEM method can be implemented without the use of MCMC and result in significant time savings in simulation studies. 


It should be noted that the global-MEM can be alternatively implemented using a constrained empirical Bayesian approach \cite{hobbs2018bayesian, kaizer2019basket, kaizer2021statistical}. The constrained empirical Bayesian implementation assumes the prior exchangeability probability is between 0 and an upper bounded $\mathbb{B}$ for baskets identified as exchangeable. The upper bound $\mathbb{B}$ helps to limit the amount of information sharing between baskets. The simulation results given by Kaizer et al.\cite{kaizer2019basket} in a 5-arm basket trial setting shows that the constrained empirical Bayesian implementation maintains the FWER at target level across all simulation scenarios. However, it leads to much smaller power in identifying promising baskets compared to the full Bayesian implementation. The addition of another tuning parameter $\mathbb{B}$ also makes the global-MEM more difficult to calibrate.

\subsection{Comparison with Simon's two-stage design}  \label{sec:simon_cmp}
In case 2, we compare the local-MEM method to Simon's two stage design, which is commonly used in phase II oncology trials. We implemented Simon's "minimax" two-stage design for each basket independently using the Bonferroni method to adjust for multiplicity. Consider a null response rate of 15\% and an alternative response rate of 45\%. The minimax two-stage design requires a maximum sample size of 16 to achieve a basket-wise power of 80\% at a basket-wise type I error rate of 0.025, and conducts an interim futility analysis when  $n_b=10$ patients have been accrued and evaluated. To facilitate comparisons, the local-MEM method is tested assuming the same maximum basket sample size ($N_b=16$) and the same interim sample size ($n_b=10$).  

\begin{figure}[ht]
\centering
\includegraphics[scale=0.8]{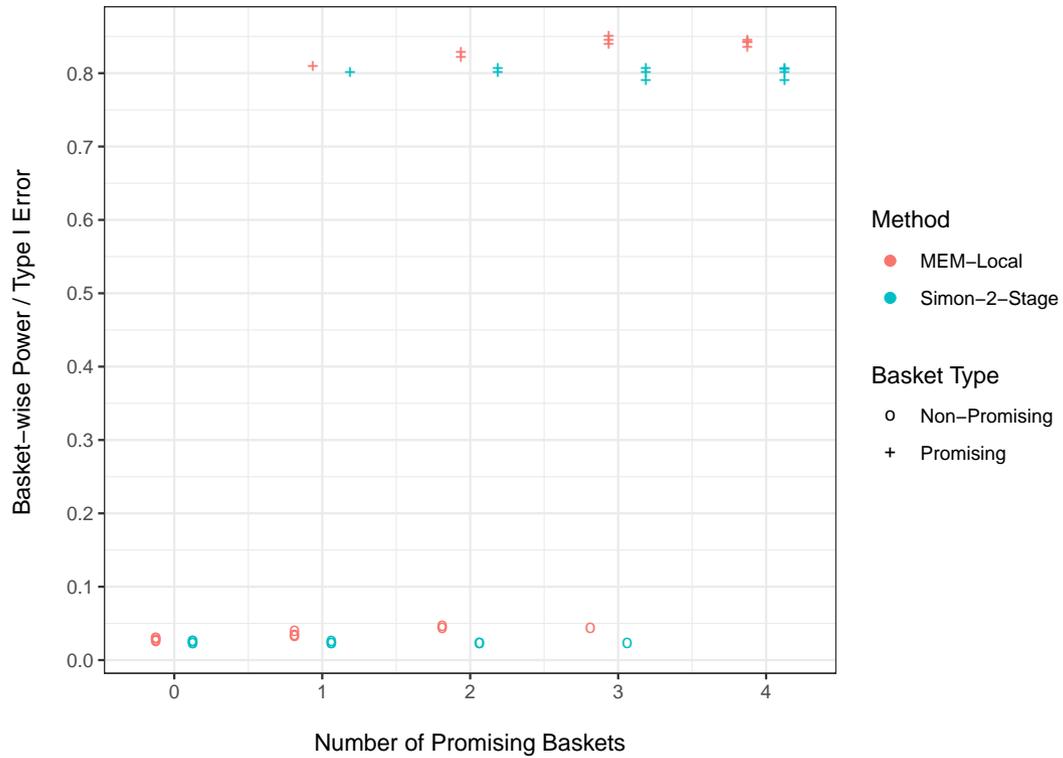}
\caption{The performance of a two-stage design based on the local-MEM method ($\delta=2$) in comparison with Simon's minimax two-stage design. FWER under the global null scenario is controlled at 0.10 level for both designs (Local-MEM: FWER $=0.098$, Simon two-stage: FWER $=0.091$). For both designs, a futility analysis is conducted when 10 patients have been accrued in each basket and the
the maximum sample size of each basket is 16.   \label{2stg_compare}}
\end{figure}

By controlling FWER under the global null (FWER$<0.10$) and maximizing trial-wise power under the global alternative, the stopping boundary is chosen to be $q_1=0.703, q_2=0.977$ when $\delta=2$. That is, any single basket needs to achieve posterior probability greater than
0.703 in stage I to continue to the next stage, and a posterior probability greater than 0.977 in stage II to be claimed as efficacious. To satisfy
the target FWER and basket-wise power, the Simon minimax two-stage design yields its cutoff as $r_1=1, r_2=5$. That is, the trial is stopped at stage I if $\le 1$ responses are seen, and the treatment is considered as promising if at least $6$ responses are observed in the 16 patients. Figure \ref{2stg_compare} shows the basket-wise power and type I error rates of the proposed local-MEM($\delta=2$) method for all simulation scenarios as compared to the Simon's two-stage design. Across all the different scenarios, the local-MEM($\delta=2$) method is able to maintain basket-wise type I error rate
for non-promising baskets at an acceptable level ($<0.05$). The local-MEM approach also provides more power under mixed alternative and the global alternative scenarios compared to the Simon two-stage design. 

\begin{figure}[ht]
\centering
\includegraphics[scale=0.9]{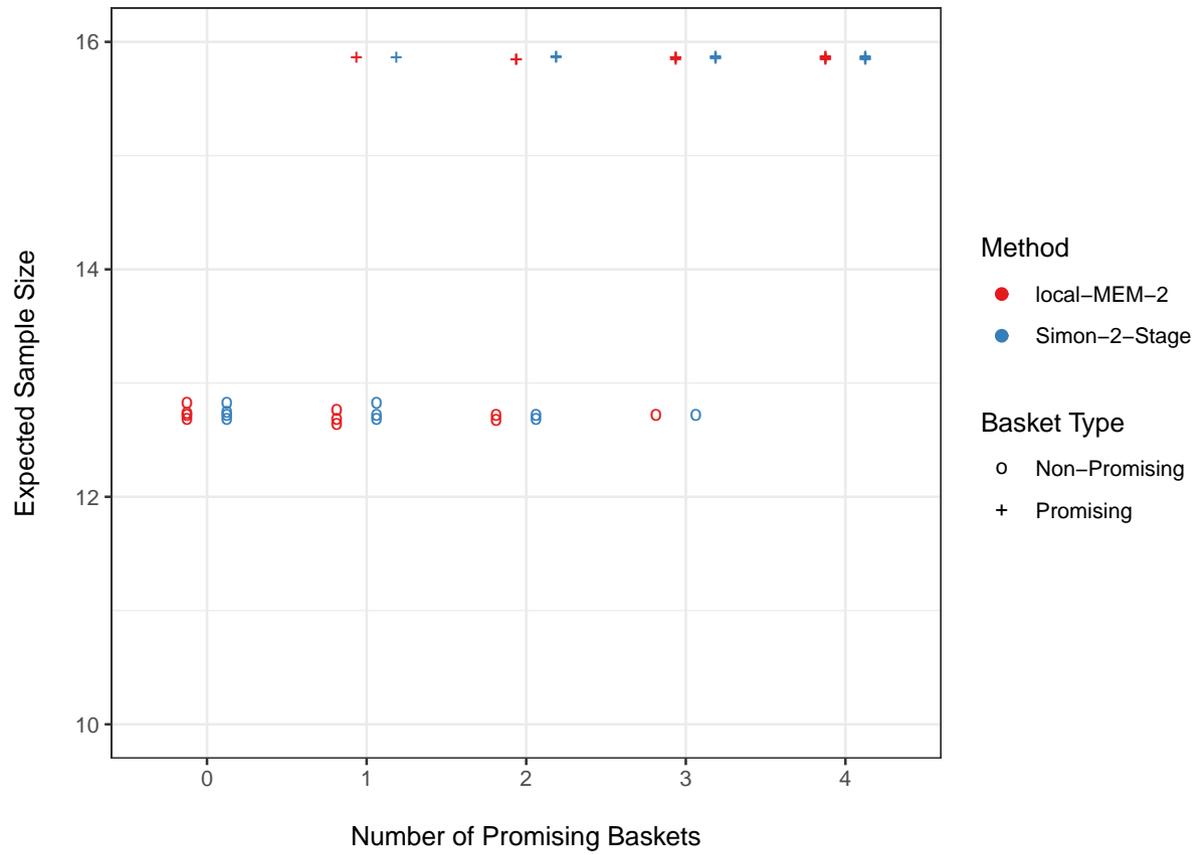}
\caption{Comparison of basket-wise expected sample size: local MEM ($\delta=2$) vs. Simon's minimax two-stage design. \label{en_plot-1}}
\end{figure}

More detailed comparisons between local-MEM and Simon's two-stage design can be found in Table \ref{tab:2stg_cmp}. As expected, Simon's two-stage design using Bonferroni adjustment maintains the FWER under the global null(0.091) and each of the mixed alternative scenarios. Similar to results from the fixed sample size case, different choices of  $\delta$ leads to different levels of FWER control for the local-MEM. With $\delta=2$, the FWER level is less than the target level of 0.10 for all the mixed alternative scenarios, whereas smaller values of $\delta$ results in inflated FWER under the "1 success" and "2 success" scenarios. Compared to Simon's minimax two-stage design, the local-MEM models with different prior choices ($\delta=\{0,1,2\}$) all have greater basket-wise power in promising baskets across different scenarios. When there's only one promising basket ("1 success"), local-MEM achieves a basket-wise power similar to Simon's two stage design. Setting $\delta=0$ leads to higher basket-wise power compared to the local-MEM with $\delta=1$ and $\delta=2$. When all baskets are promising, the basket-wise power is close to $90\%$ for $\delta=0$, $\sim 86\%$ for $\delta=1$, and $\sim 84\%$ for $\delta=2$. However, this gain in power is associated with inflated FWER, especially under the "1 success" scenario. Under the "1 success" scenario, the FWER for local-MEM with $\delta=0, 1$ is 0.131 and 0.118, respectively. The expected sample size (EN) for local-MEM($\delta=2$) and Simon two-stage minimax design are shown to be very similar (Figure \ref{en_plot-1}). For non-promising baskets, both Simon two-stage minimax design and local-MEM ($\delta=2$) have an expected sample size around 12.7. For promising baskets, both have expected sample size close to 16. 

\subsection{Sensitivity analysis}  \label{sec:sens}
We conduct a series of sensitivity analysis to evaluate the performance of the local-MEM by considering 
different prior choices ($\delta=\{0, 1, 2\}$)
and different response rates under the null and alternative scenarios. The performance of local-MEM with $\delta=\{0, 1, 2\}$ under fixed and two-stage design has been summarized in Table \ref{tab:b4_compare}, Table \ref{tab:2stg_cmp} and discussed in previous sections. We report the performance of local-MEM under more heterogeneous response scenarios in Table \ref{tab:b4_compare_hetero}.
Specifically, under the null, the probability of response for basket A, B, C, D is assumed to be 0.25, 0.25, 0.15, and 0.15, respectively. Under the alternative, we assume a $30\%$ increment in response rate. That is, the probability of response for basket A, B, C, D is assumed to be 0.55, 0.55, 0.45, and 0.45, respectively. 

Table \ref{tab:b4_compare_hetero} shows comparisons of basket-wise power / type I error rate, as well as FWER for local-MEM ($\delta=\{0,1,2\}$), global-MEM (prior exchangeability=$\{0.1, 0.5\}$), EXNEX, RoBoT, and basket-wise analysis using uniform prior for 4 baskets with equal sample size ($N=19$) for different response scenarios. The basket-wise analysis is the most conservative design among all listed methods. It only achieves at most $83.7\%$ basket-wise power for promising baskets. In contrast, all the other designs provide different levels of power gains. The global-MEM (0.5) achieves the greatest basket-wise power ($> 94.3\%$) in the "4 success" scenario, and as much as $93.8\%$ basket-wise power for "3 success" scenario. However, it's aggressive information borrowing strategy leads to notable inflation of FWER and basket-wise type I error rate when the trial contains a mixture of promising and non-promising baskets. The global-MEM method with prior exchangeability 0.1 and RoBoT have similar operating characteristics. Both methods are able to achieve higher power for promising baskets compared to basket-wise analysis and have modestly inflated FWER. Under the "1 success"  scenario, the FWER for global-MEM(0.1) and RoBoT is 0.121 and 0.115, respectively. Under the "2 success"  scenario, the respective FWER for global-MEM(0.1) and RoBoT is 0.102 and 0.117.

The FWER is well controlled for local-MEM ($\delta=\{0,1,2\}$), EXNEX and basket-wise analysis across all scenarios. The basket-wise power of local-MEM decreases as $\delta$ increases. The local-MEM($\delta=0$) provides higher power for promising baskets compared to EXNEX and has a slightly inflated FWER under the "1 success" scenario (FWER=0.103), whereas local-MEM($\delta=2$)
provides basket-wise power comparable to EXNEX and maintains FWER < 0.1 in all scenarios. Local-MEM promotes information sharing for baskets with similar response rates and result in unbalanced power gains among promising baskets. Under the "4 success" scenario, baskets A and B gain less power than baskets C and D because their estimated response rates are pooled lower towards the overall mean when all baskets appear to have similar response rates, i.e., in the same block. In contrast, the RoBoT approach provides significant power gains in all promising baskets as it borrows information according to the likelihood of the alternative hypotheses being true rather than the similarity in response rates. 

We further extend the sensitivity analysis to include 6 baskets, assuming a sample size of 19 for each basket. We assume the probability of response is $p_0=0.15$ for non-promising baskets and $p_1=0.45$ for promising baskets. We compare the performance across the aforementioned methods: local-MEM ($\delta=\{0,1,2\}$), global-MEM (0.1, 0.5), basket-wise analysis, EXNEX, and RoBoT. Table \ref{tab:b6_compare_homo} shows the operating characteristics of these methods across different scenarios. As shown in Table \ref{tab:b6_compare_homo}, the basket-wise analysis is still the most conservative design, which maintains FWER strictly controlled $< 0.1$ across all scenarios but only achieves $82.1\% \sim 83.7\%$ basket-wise power. The global-MEM models achieve the highest basket-wise power when $\ge 5$ out of 6 baskets are promising. In particular, when all baskets are promising, it obtains $\ge 98.5\%$ basket-wise power with a prior exchangeability of 0.5 and $\ge 94.3\%$ basket-wise power with a prior exchangeability of 0.1. However, as is in the 4-basket case, the global-MEM has overly inflated FWER, and the inflation is more severe here in the 6-basket case.
The local-MEM method ($\delta=\{0,1,2\}$) is able to maintain basket-wise type I error rate at an acceptable level ($<0.05$) in all scenarios and has lower FWER compared to global-MEM. Compared to RoBoT, the local-MEM with $\delta=2$ provides better control of FWER and achieves higher basket-wise power when fewer baskets are promising ($\sim 84\%$ vs. $\sim 82\%$ in the "1 success" scenario), whereas RoBoT outperforms local-MEM ($\delta=2$) in terms of basket-wise power when more baskets are promising. Similar to the 4-basket case, the EXNEX approach is able to maintain FWER in all scenarios, but its power gains only become noticeable when a large number of promising baskets are present under the "5 success" and "6 success" scenarios.

\section{Discussion}  \label{discussion}
In this article, we propose an information borrowing strategy under the local multisource exchangeability (local-MEM) assumption. Unlike the global-MEM framework, which allows information sharing among all possible pairs of baskets, we restrict information sharing to a similarity based local scope. Based on simulation studies for fixed and two-stage designs, we demonstrate the proposed local-MEM approach is effective in recognizing baskets with similar response rates and borrowing strength from these baskets. Compared with basket-wise analysis, our approach achieves greater basket-wise power while maintaining both basket-wise and family-wise type I error rates at an acceptable level. In comparison with the global-MEM method using the full Bayesian implementation \cite{hobbs2018bayesian}, our approach shows better control of family-wise type I error rate.
Moreover, our proposed approach is more computationally efficient compared with other Bayesian approaches, as all posterior quantities of interest could be derived explicitly without the need for sampling algorithms.

The majority of basket trials are designed as exploratory studies with the aim of identifying promising indications for future confirmatory studies. Recently, novel statistical designs have been developed to accommodate the need for phase III, confirmatory basket trials \cite{beckman2016adaptive, chen2016statistical}. When multiple indications are tested simultaneously, careful considerations should be made to balance the trade off between statistical power and FWER under both exploratory and confirmatory settings. According to Dmitrienko et al\cite{dmitrienko2009multiple}, there are two levels of FWER control: weak vs. strong. In the context of basket trials, weak control requires the FWER to be maintained at the target level under the global null scenario, whereas strong control requires the FWER to be guaranteed under all possible scenarios. Kaizer et al.\cite{kaizer2019basket} considered different calibrations of the MEM approach with varying levels of type I error rate control. For phase II, proof of concept basket trials, Kaizer et al.\cite{kaizer2019basket} suggested designs based on a weak control of type I error rate can be applied to maintain the type I error rate either at basket level or in the scenario when only one null basket presents. On the other hand, designs allowing for strong control of FWER is more appropriate for confirmatory basket trials leading to drug approval. As our objective is to construct statistical designs for exploratory basket trials which often have small basket sample size and unbalanced accrual, we focus on the weak control of FWER and calibrate the proposed design to maximize the trial-wise power. 

The choice of prior in our approach can be easily adapted for different circumstances in practice. The presence of patient-level information, such as data from high throughout sequencing analysis might help to predict treatment benefit in different baskets. When such information is available, we can assign a more informative prior distribution to the different partitions of baskets by assuming some partitions are more likely to respond similarly than others. Likewise, we can consider some partitions as unlikely based on biology of the disease. 

Some limitations of our approach should also be noted. The local-MEM framework enumerates all possible partitions and evaluate the uncertainties related to each partition. This process can become computationally intensive when the number of baskets is large. Alternative methods based on clustering or mixture models can be applied when there is a large number of baskets. \cite{zhou2020robot,kang2021hierarchical, neuenschwander2016robust} 


\section*{Acknowledgments}
We thank Dr. Alexander M. Kaizer for the very helpful comments and suggestions on this paper. 



\subsection*{Financial disclosure}

None reported.

\subsection*{Conflict of interest}

The authors declare no potential conflict of interests.

\newpage
\bibliography{main}%

\section*{Supporting information}

The following supporting information is available as part of the online article.


\newpage


\newpage
\begin{table}[ht]
\centering
\caption{Partitions of four baskets with different prior choices ($\delta=\{0, 1, 2\}$). The number of blocks in the $j$-th partition is denoted as $K_j$.}
\label{tab:pp}
\scalebox{0.9}{
\begin{tabular}{cccccccc}
  \toprule
\multicolumn{4}{c}{Membership} &  & \multicolumn{3}{c}{Prior Choice}  \\
  A & B & C & D & $K_{j}$ &$\delta=0$ & $\delta=1$ & $\delta=2$ \\
  \midrule
  1 & 1 & 1 & 1 &    1 & 0.067 & 0.027 & 0.010 \\ 
  1 & 1 & 1 & 2 &    2 & 0.067 & 0.054 & 0.040 \\ 
  1 & 1 & 2 & 1 &    2 & 0.067 & 0.054 & 0.040 \\ 
  1 & 2 & 1 & 1 &    2 & 0.067 & 0.054 & 0.040 \\ 
  2 & 1 & 1 & 1 &    2 & 0.067 & 0.054 & 0.040 \\ 
  1 & 2 & 2 & 1 &    2 & 0.067 & 0.054 & 0.040 \\ 
  1 & 1 & 2 & 2 &    2 & 0.067 & 0.054 & 0.040 \\ 
  1 & 2 & 1 & 2 &    2 & 0.067 & 0.054 & 0.040 \\ 
  1 & 2 & 3 & 1 &    3 & 0.067 & 0.081 & 0.091 \\ 
  1 & 1 & 2 & 3 &    3 & 0.067 & 0.081 & 0.091 \\ 
  1 & 2 & 1 & 3 &    3 & 0.067 & 0.081 & 0.091 \\ 
  2 & 1 & 3 & 1 &    3 & 0.067 & 0.081 & 0.091 \\ 
  2 & 1 & 1 & 3 &    3 & 0.067 & 0.081 & 0.091 \\ 
  2 & 3 & 1 & 1 &    3 & 0.067 & 0.081 & 0.091 \\ 
  1 & 2 & 3 & 4 &    4 & 0.067 & 0.108 & 0.162 \\ 
   \bottomrule
\end{tabular}
}
\end{table}

\newpage

\begin{table}[ht]
\centering
\caption{Operating characteristics of local-MEM using different prior choices ($\delta=\{0,1,2\}$), EXNEX analysis, global-MEM (prior exchangeability=0.1, 0.5), basket-wise analysis using uniform prior, and RoBoT. The null and alternative response rate for each basket is shown in parenthesis. Promising baskets are in bold format. FWER under the global null scenario is controlled at 0.10 level for each method.}
\label{tab:b4_compare}
\scalebox{0.85}{
\begin{tabular}{cccccc|ccccc}
  \toprule
 & A & B & C & D & \multirow{2}{*}{FWER} & A & B & C & D & \multirow{2}{*}{FWER} \\
 & (0.15/0.45) & (0.15/0.45) & (0.15/0.45) & (0.15/0.45) &  & (0.15/0.45) & (0.15/0.45) & (0.15/0.45) & (0.15/0.45) &  \\
  \hline
  & \multicolumn{5}{c|}{Local MEM, $\delta=0$} & \multicolumn{5}{c}{EXNEX}\\
  \hline
  0 Success & 0.026 & 0.027 & 0.025 & 0.022 & 0.085 & 0.018 & 0.016 & 0.015 & 0.016 & 0.063 \\ 
  1 Success & 0.048 & 0.052 & 0.049 & \textbf{0.853} & 0.134 & 0.014 & 0.012 & 0.015 & \textbf{0.825} & 0.040 \\ 
  2 Success & 0.063 & 0.066 & \textbf{0.904} & \textbf{0.896} & 0.116 & 0.021 & 0.015 & \textbf{0.825} & \textbf{0.828} & 0.034 \\ 
  3 Success & 0.081 & \textbf{0.907} & \textbf{0.914} & \textbf{0.906} & 0.081 & 0.031 & \textbf{0.835} & \textbf{0.830} & \textbf{0.831} & 0.031 \\ 
  4 Success & \textbf{0.933} & \textbf{0.927} & \textbf{0.934} & \textbf{0.929} & - & \textbf{0.861} & \textbf{0.852} & \textbf{0.852} & \textbf{0.854} & - \\
  \hline
  & \multicolumn{5}{c|}{Local MEM, $\delta=1$} & \multicolumn{5}{c}{Global MEM, 0.5}\\
  \hline
  0 Success & 0.028 & 0.031 & 0.027 & 0.026 & 0.097 & 0.009 & 0.007 & 0.061 & 0.057 & 0.096 \\ 
  1 Success & 0.042 & 0.044 & 0.042 & \textbf{0.861} & 0.117 &   0.027 & 0.027 & 0.144 & \textbf{0.808} & 0.179 \\ 
  2 Success & 0.051 & 0.056 & \textbf{0.898} & \textbf{0.890} & 0.104 &   0.110 & 0.109 & \textbf{0.924} & \textbf{0.920} & 0.151 \\ 
  3 Success & 0.053 & \textbf{0.900} & \textbf{0.908} & \textbf{0.900} & 0.053 &   0.218 & \textbf{0.855} & \textbf{0.938} & \textbf{0.936} & 0.218 \\ 
  4 Success & \textbf{0.909} & \textbf{0.903} & \textbf{0.911} & \textbf{0.905} & - &   \textbf{0.946} & \textbf{0.952} & \textbf{0.943} & \textbf{0.947} & - \\  
  \hline
  & \multicolumn{5}{c|}{Local MEM, $\delta=2$} & \multicolumn{5}{c}{Global MEM, 0.1}\\
  \hline
  0 Success & 0.028 & 0.030 & 0.027 & 0.026 & 0.097 & 0.034 & 0.031 & 0.034 & 0.028 & 0.099 \\  
  1 Success & 0.034 & 0.036 & 0.033 & \textbf{0.860} & 0.096 &   0.056 & 0.053 & 0.059 & \textbf{0.865} & 0.136 \\  
  2 Success & 0.043 & 0.043 & \textbf{0.876} & \textbf{0.871} & 0.083 &   0.080 & 0.074 & \textbf{0.903} & \textbf{0.907} & 0.135 \\  
  3 Success & 0.043 & \textbf{0.888} & \textbf{0.897} & \textbf{0.886} & 0.043 &   0.109 & \textbf{0.931} & \textbf{0.930} & \textbf{0.934} & 0.109 \\  
  4 Success & \textbf{0.897} & \textbf{0.888} & \textbf{0.897} & \textbf{0.885} & -  &   \textbf{0.947} & \textbf{0.949} & \textbf{0.948} & \textbf{0.948} & - \\ 
  \hline
  & \multicolumn{5}{c|}{Basketwise analysis} & \multicolumn{5}{c}{RoBoT}\\
  \hline
  0 Success& 0.017 & 0.018 & 0.015 & 0.015 & 0.063 & 0.027 & 0.022 & 0.027 & 0.029 & 0.098 \\
  1 Success& 0.017 & 0.018 & 0.015 & \textbf{0.827} & 0.049 & 0.048 & 0.034 & 0.046 & \textbf{0.858} & 0.124 \\ 
  2 Success& 0.017 & 0.018 & \textbf{0.837} & \textbf{0.827} & 0.034 & 0.061 & 0.049 & \textbf{0.877} & \textbf{0.886} & 0.106 \\
  3 Success&  0.017 & \textbf{0.821} & \textbf{0.837} & \textbf{0.827} & 0.017 & 0.062 & \textbf{0.921} & \textbf{0.920} & \textbf{0.916} & 0.062 \\ 
  4 Success& \textbf{0.830} & \textbf{0.821} & \textbf{0.837} & \textbf{0.827} & - & \textbf{0.916} & \textbf{0.926} & \textbf{0.921} & \textbf{0.920} & - \\ 
   \bottomrule
\end{tabular}
}
\end{table}

\newpage

\begin{table}[ht]
\centering
\caption{Operating characteristics of the two-stage local-MEM design  ($\delta=\left\{0, 1, 2\right\}$) and Simon's minimax two-stage design. For both methods, the maximum sample size of each basket is 16 and interim analysis is conducted when 10 patients have been accrued and evaluated. The null and alternative response rate
for each basket is shown in parenthesis. Promising baskets are in bold format. FWER under the global null scenario is controlled at 0.10 level for each method.}
\label{tab:2stg_cmp}
\scalebox{0.95}{
\begin{tabular}{cccccc}
  \toprule
   & A & B & C & D & \multirow{2}{*}{FWER}\\
 & (0.15/0.45) & (0.15/0.45) & (0.15/0.45) & (0.15/0.45) \\\hline
    \multicolumn{6}{c}{Local MEM, $\delta=0$}\\\hline
0 Success & 0.028 & 0.029 & 0.032 & 0.032 & 0.099 \\ 
  1 Success & 0.046 & 0.048 & 0.051 & \textbf{0.801} & 0.131 \\ 
  2 Success & 0.060 & 0.059 & \textbf{0.843} & \textbf{0.835} & 0.110 \\ 
  3 Success & 0.098 & \textbf{0.855} & \textbf{0.865} & \textbf{0.854} & 0.098 \\ 
  4 Success & \textbf{0.908} & \textbf{0.895} & \textbf{0.904} & \textbf{0.902} & - \\\hline
  \multicolumn{6}{c}{Local MEM, $\delta=1$}\\\hline
  0 Success & 0.027 & 0.028 & 0.032 & 0.032 & 0.100 \\ 
  1 Success & 0.040 & 0.043 & 0.046 & \textbf{0.805} & 0.118 \\ 
  2 Success & 0.048 & 0.051 & \textbf{0.836} & \textbf{0.826} & 0.096 \\ 
  3 Success & 0.055 & \textbf{0.842} & \textbf{0.854} & \textbf{0.844} & 0.055 \\ 
  4 Success & \textbf{0.865} & \textbf{0.856} & \textbf{0.865} & \textbf{0.861} & - \\ \hline
  \multicolumn{6}{c}{Local MEM, $\delta=2$}\\\hline
  0 Success & 0.026 & 0.028 & 0.031 & 0.030 & 0.098 \\ 
  1 Success & 0.033 & 0.035 & 0.039 & \textbf{0.810} & 0.098 \\ 
  2 Success & 0.044 & 0.047 & \textbf{0.829} & \textbf{0.822} & 0.089 \\ 
  3 Success & 0.044 & \textbf{0.839} & \textbf{0.850} & \textbf{0.845} & 0.044 \\ 
  4 Success & \textbf{0.843} & \textbf{0.835} & \textbf{0.845} & \textbf{0.842} & - \\ \hline
  \multicolumn{6}{c}{Simon two-stage ($r_1=1, r_2=5$)}\\\hline
  0 Success & 0.023 & 0.024 & 0.023 & 0.024 & 0.091 \\ 
  1 Success & 0.024 & 0.023 & 0.024 & \textbf{0.804} & 0.069 \\ 
  2 Success & 0.024 & 0.025 & \textbf{0.794} & \textbf{0.792} & 0.050 \\ 
  3 Success & 0.025 & \textbf{0.797} & \textbf{0.806} & \textbf{0.804} & 0.025 \\ 
  4 Success & \textbf{0.803} & \textbf{0.798} & \textbf{0.802} & \textbf{0.797} & - \\ 
   \bottomrule
\end{tabular}
}
\end{table}

\newpage

\begin{table}[ht]
\centering
\caption{Operating characteristics of local-MEM using different prior choices ($\delta=\{0,1,2\}$), EXNEX analysis, global MEM (prior exchangeability=0.1, 0.5), basket-wise analysis, and RoBoT. The null and alternative response rate of each basket is shown in parenthesis. Promising baskets are in bold format.}
\label{tab:b4_compare_hetero}
\scalebox{0.85}{
\begin{tabular}{cccccc|ccccc}
  \toprule
 & A & B & C & D & FWER & A & B & C & D & FWER \\ 
 & (0.25/0.55) & (0.25/0.55) & (0.15/0.45) & (0.15/0.45) &  & (0.25/0.55) & (0.25/0.55) & (0.15/0.45) & (0.15/0.45) &  \\ 
  \hline
  & \multicolumn{5}{c}{Local MEM, $\delta=0$} & \multicolumn{5}{c}{EXNEX}\\
  \hline
  0 Success & 0.013 & 0.011 & 0.030 & 0.027 & 0.076 & 0.020 & 0.016 & 0.019 & 0.019 & 0.072 \\ 
  1 Success & 0.030 & 0.032 & 0.046 & \textbf{0.863} & 0.103 &   0.024 & 0.022 & 0.020 & \textbf{0.826} & 0.063 \\ 
  2 Success & 0.038 & 0.040 & \textbf{0.897} & \textbf{0.889} & 0.074 &   0.027 & 0.027 & \textbf{0.851} & \textbf{0.845} & 0.053 \\ 
  3 Success & 0.055 & \textbf{0.823} & \textbf{0.890} & \textbf{0.888} & 0.055 & 0.032 & \textbf{0.816} & \textbf{0.849} & \textbf{0.843} & 0.032 \\ 
  4 Success & \textbf{0.860} & \textbf{0.860} & \textbf{0.899} & \textbf{0.894} & - &   \textbf{0.808} & \textbf{0.820} & \textbf{0.871} & \textbf{0.862} & - \\   
  \hline
  & \multicolumn{5}{c}{Local MEM, $\delta=1$} & \multicolumn{5}{c}{Global MEM, 0.5}\\
  \hline
  0 Success & 0.012 & 0.011 & 0.025 & 0.023 & 0.067 & 0.009 & 0.007 & 0.061 & 0.057 & 0.096 \\ 
  1 Success & 0.026 & 0.028 & 0.032 & \textbf{0.852} & 0.082 & 0.027 & 0.027 & 0.144 & \textbf{0.808} & 0.179 \\ 
  2 Success & 0.031 & 0.034 & \textbf{0.870} & \textbf{0.868} & 0.065 & 0.110 & 0.109 & \textbf{0.924} & \textbf{0.920} & 0.151 \\ 
  3 Success & 0.030 & \textbf{0.816} & \textbf{0.873} & \textbf{0.870} & 0.030 & 0.218 & \textbf{0.855} & \textbf{0.938} & \textbf{0.936} & 0.218 \\ 
  4 Success & \textbf{0.819} & \textbf{0.811} & \textbf{0.880} & \textbf{0.874} & - & \textbf{0.946} & \textbf{0.952} & \textbf{0.943} & \textbf{0.947} & - \\ 
  \hline
  & \multicolumn{5}{c}{Local MEM, $\delta=2$} & \multicolumn{5}{c}{Global MEM, 0.1}\\
  \hline
  0 Success & 0.029 & 0.029 & 0.023 & 0.021 & 0.098 & 0.016 & 0.016 & 0.042 & 0.038 & 0.099 \\ 
  1 Success & 0.030 & 0.030 & 0.026 & \textbf{0.844} & 0.084 & 0.030 & 0.030 & 0.068 & \textbf{0.875} & 0.121 \\ 
  2 Success & 0.030 & 0.031 & \textbf{0.858} & \textbf{0.852} & 0.060 & 0.058 & 0.061 & \textbf{0.910} & \textbf{0.908} & 0.102 \\ 
  3 Success & 0.029 & \textbf{0.816} & \textbf{0.864} & \textbf{0.859} & 0.029 & 0.091 & \textbf{0.857} & \textbf{0.909} & \textbf{0.908} & 0.091 \\ 
  4 Success & \textbf{0.822} & \textbf{0.816} & \textbf{0.857} & \textbf{0.852} & - & \textbf{0.896} & \textbf{0.908} & \textbf{0.911} & \textbf{0.913} & - \\ 
   \hline
   & \multicolumn{5}{c}{Basketwise analysis} & \multicolumn{5}{c}{RoBoT}\\
   \hline
0 Success & 0.029 & 0.029 & 0.015 & 0.015 & 0.086 & 0.031 & 0.019 & 0.022 & 0.029 & 0.096 \\ 
  1 Success & 0.029 & 0.029 & 0.015 & \textbf{0.827} & 0.073 & 0.047 & 0.034 & 0.042 & \textbf{0.855} & 0.115 \\ 
  2 Success & 0.029 & 0.029 & \textbf{0.837} & \textbf{0.827} & 0.058 & 0.068 & 0.057 & \textbf{0.878} & \textbf{0.887} & 0.117 \\ 
  3 Success & 0.029 & \textbf{0.815} & \textbf{0.837} & \textbf{0.827} & 0.029 & 0.087 & \textbf{0.890} & \textbf{0.918} & \textbf{0.915} & 0.087 \\ 
  4 Success & \textbf{0.821} & \textbf{0.815} & \textbf{0.837} & \textbf{0.827} & - & \textbf{0.902} & \textbf{0.922} & \textbf{0.921} & \textbf{0.920} & - \\ 
   \bottomrule
\end{tabular}
}
\end{table}
\newpage

\begin{table}[ht]
\centering\caption{Operating characteristics of local-MEM and other approaches when six baskets are studied. The probability of response is 0.15(0.45) for all baskets. Promising baskets are in bold format across scenarios.}
\label{tab:b6_compare_homo}
\scalebox{0.9}{
\begin{tabular}{cccccccc|ccccccc}
  \toprule
  & A & B & C & D & E & F & FWER & A & B & C & D & E & F & FWER\\ 
  \midrule
  & \multicolumn{7}{c}{Local MEM, $\delta=0$} & \multicolumn{7}{c}{EXNEX}\\
  \hline
  0 Success & 0.019 & 0.020 & 0.019 & 0.017 & 0.020 & 0.018 & 0.096 & 0.011 & 0.015 & 0.017 & 0.017 & 0.012 & 0.011 & 0.080 \\ 
  1 Success & 0.035 & 0.035 & 0.034 & 0.030 & 0.036 & \textbf{0.839} & 0.152 & 0.020 & 0.019 & 0.021 & 0.012 & 0.012 & \textbf{0.815} & 0.078 \\ 
  2 Success & 0.040 & 0.041 & 0.040 & 0.038 & \textbf{0.879} & \textbf{0.879} & 0.145 & 0.016 & 0.018 & 0.021 & 0.014 & \textbf{0.833} & \textbf{0.832} & 0.069 \\ 
  3 Success & 0.039 & 0.041 & 0.040 & \textbf{0.883} & \textbf{0.893} & \textbf{0.896} & 0.115 & 0.028 & 0.016 & 0.021 & \textbf{0.830} & \textbf{0.822} & \textbf{0.845} & 0.063 \\ 
  4 Success & 0.038 & 0.039 & \textbf{0.893} & \textbf{0.883} & \textbf{0.890} & \textbf{0.891} & 0.075 & 0.030 & 0.032 & \textbf{0.856} & \textbf{0.844} & \textbf{0.845} & \textbf{0.839} & 0.061 \\ 
  5 Success & 0.044 & \textbf{0.878} & \textbf{0.888} & \textbf{0.884} & \textbf{0.890} & \textbf{0.890} & 0.044 & 0.042 & \textbf{0.859} & \textbf{0.867} & \textbf{0.870} & \textbf{0.849} & \textbf{0.868} & 0.042 \\ 
  6 Success & \textbf{0.902} & \textbf{0.891} & \textbf{0.898} & \textbf{0.894} & \textbf{0.900} & \textbf{0.901} & - & \textbf{0.911} & \textbf{0.915} & \textbf{0.895} & \textbf{0.920} & \textbf{0.894} & \textbf{0.904} & - \\ 
  \hline
  & \multicolumn{7}{c}{Local MEM, $\delta=1$} & \multicolumn{7}{c}{Global MEM, 0.5}\\
  \hline
  0 Success & 0.019 & 0.020 & 0.019 & 0.017 & 0.020 & 0.018 & 0.095 & 0.045 & 0.047 & 0.040 & 0.045 & 0.051 & 0.041 & 0.098 \\ 
  1 Success & 0.032 & 0.031 & 0.031 & 0.029 & 0.032 & \textbf{0.839} & 0.137 & 0.099 & 0.098 & 0.100 & 0.101 & 0.096 & \textbf{0.791} & 0.177 \\ 
  2 Success & 0.038 & 0.038 & 0.035 & 0.035 & \textbf{0.871} & \textbf{0.870} & 0.134 & 0.179 & 0.174 & 0.196 & 0.181 & \textbf{0.878} & \textbf{0.879} & 0.325 \\ 
  3 Success & 0.038 & 0.038 & 0.037 & \textbf{0.877} & \textbf{0.884} & \textbf{0.884} & 0.108 & 0.256 & 0.235 & 0.246 & \textbf{0.928} & \textbf{0.939} & \textbf{0.926} & 0.408 \\ 
  4 Success & 0.037 & 0.037 & \textbf{0.889} & \textbf{0.878} & \textbf{0.886} & \textbf{0.887} & 0.073 & 0.296 & 0.287 & \textbf{0.975} & \textbf{0.968} & \textbf{0.972} & \textbf{0.954} & 0.417 \\ 
  5 Success & 0.038 & \textbf{0.872} & \textbf{0.885} & \textbf{0.880} & \textbf{0.888} & \textbf{0.886} & 0.038 & 0.360 & \textbf{0.986} & \textbf{0.986} & \textbf{0.982} & \textbf{0.988} & \textbf{0.975} & 0.360 \\ 
  6 Success & \textbf{0.887} & \textbf{0.872} & \textbf{0.884} & \textbf{0.878} & \textbf{0.885} & \textbf{0.885} & - & \textbf{0.985} & \textbf{0.990} & \textbf{0.988} & \textbf{0.990} & \textbf{0.990} & \textbf{0.987} & - \\ 
  \hline
  & \multicolumn{7}{c}{Local MEM, $\delta=2$} & \multicolumn{7}{c}{Global MEM, 0.1}\\
  \hline
  0 Success & 0.019 & 0.020 & 0.017 & 0.017 & 0.020 & 0.018 & 0.096 & 0.020 & 0.020 & 0.020 & 0.019 & 0.021 & 0.020 & 0.096 \\ 
  1 Success & 0.025 & 0.027 & 0.024 & 0.023 & 0.027 & \textbf{0.838} & 0.108 & 0.029 & 0.029 & 0.029 & 0.028 & 0.029 & \textbf{0.816} & 0.111 \\ 
  2 Success & 0.034 & 0.035 & 0.033 & 0.033 & \textbf{0.853} & \textbf{0.856} & 0.125 & 0.048 & 0.043 & 0.047 & 0.045 & \textbf{0.847} & \textbf{0.851} & 0.140 \\ 
  3 Success & 0.036 & 0.039 & 0.035 & \textbf{0.874} & \textbf{0.878} & \textbf{0.875} & 0.105 & 0.069 & 0.063 & 0.072 & \textbf{0.877} & \textbf{0.881} & \textbf{0.881} & 0.161 \\ 
  4 Success & 0.037 & 0.037 & \textbf{0.886} & \textbf{0.879} & \textbf{0.887} & \textbf{0.880} & 0.073 & 0.087 & 0.086 & \textbf{0.910} & \textbf{0.916} & \textbf{0.912} & \textbf{0.908} & 0.152 \\ 
  5 Success & 0.034 & \textbf{0.871} & \textbf{0.884} & \textbf{0.879} & \textbf{0.889} & \textbf{0.884} & 0.034 & 0.111 & \textbf{0.940} & \textbf{0.931} & \textbf{0.937} & \textbf{0.932} & \textbf{0.934} & 0.111 \\ 
  6 Success & \textbf{0.875} & \textbf{0.860} & \textbf{0.875} & \textbf{0.868} & \textbf{0.873} & \textbf{0.874} & - & \textbf{0.946} & \textbf{0.954} & \textbf{0.943} & \textbf{0.949} & \textbf{0.948} & \textbf{0.946} & - \\ 
  \hline
  & \multicolumn{7}{c}{Basket-wise analysis} & \multicolumn{7}{c}{RoBoT}\\
  \hline
  0 success & 0.017 & 0.018 & 0.015 & 0.015 & 0.018 & 0.014 & 0.093 & 0.018 & 0.019 & 0.018 & 0.015 & 0.020 & 0.015 & 0.096 \\ 
  1 success & 0.017 & 0.018 & 0.015 & 0.015 & 0.018 & \textbf{0.831} & 0.080 & 0.027 & 0.026 & 0.025 & 0.023 & 0.025 & \textbf{0.820} & 0.113 \\ 
  2 success & 0.017 & 0.018 & 0.015 & 0.015 & \textbf{0.828} & \textbf{0.831} & 0.063 & 0.037 & 0.039 & 0.035 & 0.032 & \textbf{0.848} & \textbf{0.849} & 0.132 \\ 
  3 success & 0.017 & 0.018 & 0.015 & \textbf{0.827} & \textbf{0.828} & \textbf{0.831} & 0.049 & 0.048 & 0.050 & 0.047 & \textbf{0.877} & \textbf{0.875} & \textbf{0.874} & 0.138 \\ 
  4 success & 0.017 & 0.018 & \textbf{0.837} & \textbf{0.827} & \textbf{0.828} & \textbf{0.831} & 0.034 & 0.054 & 0.056 & \textbf{0.914} & \textbf{0.903} & \textbf{0.911} & \textbf{0.909} & 0.106 \\ 
  5 success& 0.017 &\textbf{ 0.82}1 & \textbf{0.837} & \textbf{0.827} & \textbf{0.828} & \textbf{0.831} & 0.017 & 0.056 & \textbf{0.920} & \textbf{0.925} & \textbf{0.918} & \textbf{0.926} & \textbf{0.925} & 0.056 \\ 
  6 success& \textbf{0.830} & \textbf{0.821} & \textbf{0.837} & \textbf{0.827} & \textbf{0.828} & \textbf{0.831} &- & \textbf{0.927} & \textbf{0.921} & \textbf{0.926} & \textbf{0.919} & \textbf{0.926} & \textbf{0.926} & - \\ 
   \bottomrule
\end{tabular}}
\end{table}




\newpage 

\newpage

\newpage 

\newpage 

\newpage
\begin{figure}[ht]
\centering
\includegraphics[scale=0.8]{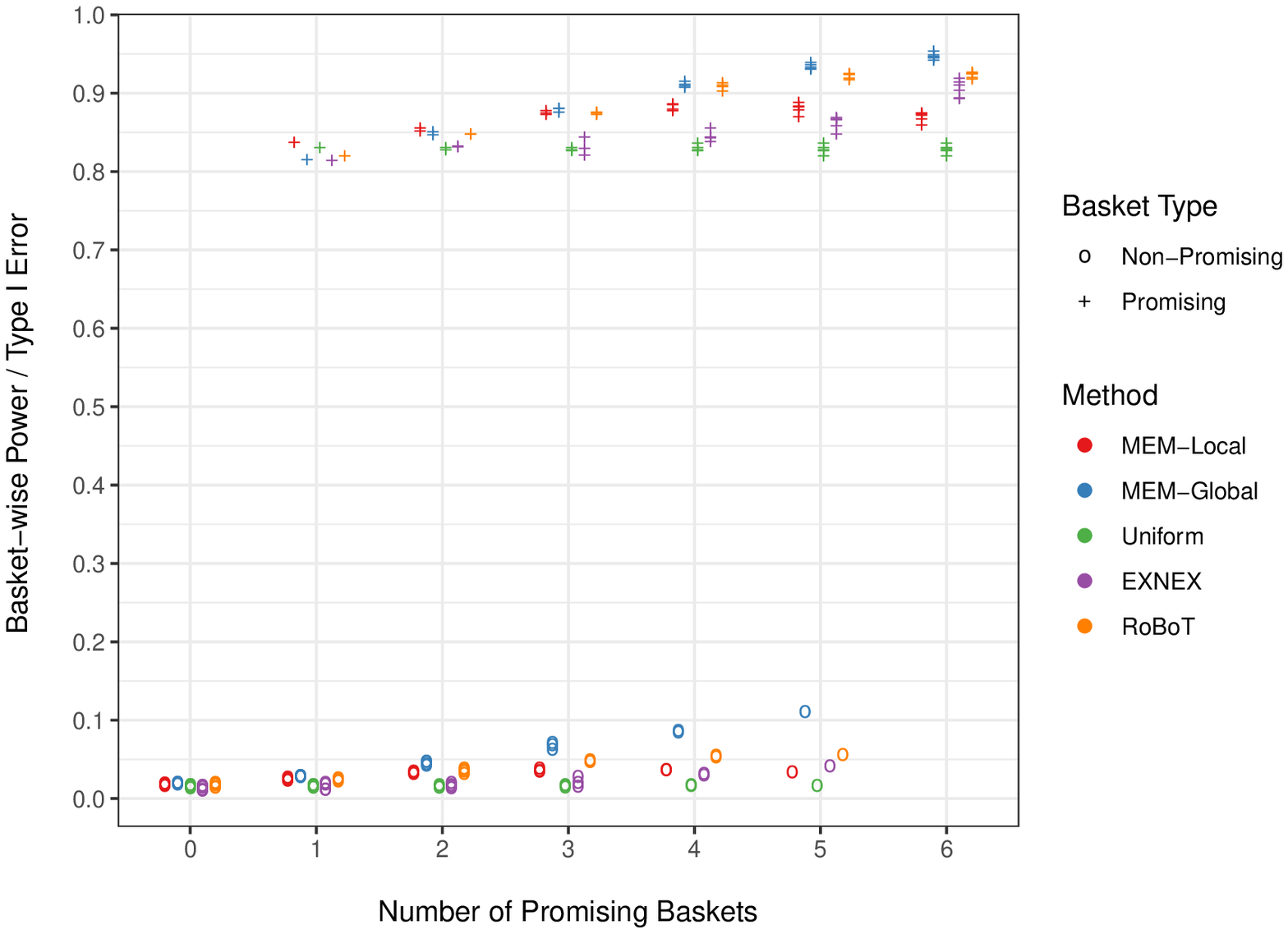}
\caption{Sensitivity analysis with a 6-basket trial. Local-MEM ($\delta=2$) vs. global MEM (prior exchangeability $=0.1$) vs. basket-wise analysis with uniform prior vs. EXNEX vs. RoBoT for studies with fixed design. Each basket has a fixed sample size of 19 without provisions for early stopping. Different methods are color coded. FWER under the global null scenario is controlled at 0.10 level for each method. \label{B6_homo_cmp}}
\end{figure}








\end{document}